\documentclass[showpacs,showkeys,twoside,eqsecnum,prd,twocolumn,aps]{revtex4-1}
\usepackage{amsfonts,amsmath,amssymb,bm,graphicx}

\begin{document}


\title{Spin-flavor oscillations of Dirac neutrinos described by
relativistic quantum mechanics}
\author{Maxim Dvornikov}
\email{maxim.dvornikov@usp.br}
\affiliation{N.~V.~Pushkov Institute of Terrestrial Magnetism,\\
Ionosphere and Radiowave Propagation (IZMIRAN),\\
142190, Troitsk, Moscow region, Russia;\\
Institute of Physics, University of S\~ao Paulo,\\
CP 66318, CEP 05315-970 S\~ao Paulo, SP, Brazil}

\date{\today}

\begin{abstract}
Spin-flavor oscillations of Dirac neutrinos in matter and a magnetic field are studied using the method of relativistic quantum mechanics. Using the exact solution of the wave equation for a massive neutrino, taking into account external fields, the effective Hamiltonian governing neutrino spin-flavor oscillations is derived. Then the The consistency of our approach with the commonly used quantum mechanical method is demonstrated. The obtained correction to the usual effective Hamiltonian results in the appearance of the new resonance in neutrino oscillations. Applications to spin-flavor neutrino oscillations in an expanding envelope of a supernova are discussed. In particular, transitions between right-polarized electron neutrinos and additional sterile neutrinos are studied for realistic background matter and magnetic field distributions. The influence of other factors such as the longitudinal magnetic field, the matter polarization, and the 
non-standard contributions to the neutrino effective potential, is also analyzed.
\end{abstract}



\maketitle

\section{Introduction}

It was confirmed by numerous experiments that neutrinos are
massive particles and there is a mixing between different neutrino
generations. Besides these non-standard model neutrino properties
it is believed that neutrinos can also have non-zero diagonal and
transition magnetic moments. The latter mix both eigenstates with
opposite helicities as well as different neutrino flavors. Thus
the existence of transition magnetic moments implies that in
presence of an external electromagnetic field conversions of
left-handed neutrinos into their right-handed counterparts of
another flavor (neutrino spin-flavor
oscillations~\cite{VolVysOku86,LimMar88}) are possible,
$\nu_\alpha^\mathrm{-{}} \leftrightarrow \nu_\beta^\mathrm{+{}}$,
where $\alpha \neq \beta$ and the index $\pm{}$ corresponds to
different neutrino helicities.

In this work we will consider spin-flavor oscillations of Dirac
neutrinos in dense matter under the influence of strong external
magnetic fields. Although the existence of Majorana neutrinos is
favored in various neutrino mass generation models, like see-saw
mechanism~\cite{FukYan03p391}, the question whether neutrinos are
Dirac or Majorana particles is still open~\cite{AviEllEng08}.

Note that Dirac and Majorana neutrinos have completely different
structure of the magnetic moments matrix. Dirac neutrinos can have
both diagonal and transition magnetic moments whereas Majorana
neutrinos only transition ones~(see, e.g.,
Ref.~\cite{FukYan03p461}). Moreover the magnetic moments matrix is
symmetric in the Dirac case and anti-symmetric for Majorana
neutrinos~\cite{FukYan03p461}.


The strongest laboratory constraint on the neutrino effective
magnetic moment, obtained by the GEMMA
collaboration~\cite{nuMMexp}, is $3.2 \times 10^{-11}
\mu_\mathrm{B}$, where $\mu_\mathrm{B}$ is the Bohr magneton.
Slightly weaker upper bound on the magnetic moments of neutrinos
was obtained by the BOREXINO collaboration~\cite{Arp08}. Note that
the astrophysical limits on the magnetic moments of Dirac and
Majorana neutrinos extracted from the white dwarfs cooling
rate~\cite{Raf99} can be even stronger than the laboratory ones.
Therefore in most realistic situations, where the magnetic field
is not too strong, neutrino spin and spin-flavor oscillations are
likely to play a sub-leading role~\cite{AkhPul03}. Nevertheless in
some astrophysical media the dynamics of neutrino oscillations can
be significantly affected by a magnetic field because of its
interaction with neutrino magnetic moments. For example, during a
supernova explosion or in the vicinity of a neutron star, magnetic
fields can reach values up to
$10^{16}\thinspace\mathrm{G}$~\cite{Aki03}, which is enough to
influence the neutrino oscillations process in a significant way.

Neutrino spin and spin-flavor oscillations in a supernova were
discussed in Refs.~\cite{LimMar88,Vol88}. The impact of neutrino
magnetic moments and spin-flavor oscillations on $r$-process
nucleosynthesis during a supernova explosion was studied in
Ref.~\cite{BalVolWel07}. The neutrino spin flip, i.e. the
transformation like $\nu_\alpha^\mathrm{-{}} \leftrightarrow
\nu_\alpha^\mathrm{+{}}$ within the same flavor, can happen not
only because of the magnetic moment interaction with an external
magnetic field, but also in collisions with background matter
during a supernova explosion, that was examined in
Ref.~\cite{Not88}.

In this work we study neutrino spin-flavor oscillations in frames
of the relativistic quantum
mechanics~\cite{Dvo,Dvo06,DvoMaa07,Dvo08JPCS,Dvo08JPG}. Within the
developed formalism we find the wave functions of neutrino mass
eigenstates, which are the superposition of flavor neutrinos, for
a given initial condition (Sec.~\ref{DENSEMATTSTRONGBQFT}). Using
this method for the description of the neutrino evolution one can
exactly take into account neutrino masses and the influence of
external fields since we use exact solutions of the Dirac equation
for massive Dirac neutrinos. Note that neutrino oscillations in
dense matter and strong magnetic fields can be also described with
help of the methods of finite temperature field
theory~\cite{NotRaf88}.

Then, in Sec.~\ref{QM}, we analyze the conventional quantum
mechanical approach to the description of neutrino spin-flavor
oscillations~\cite{LimMar88}. We discuss the cases of the neutrino
propagation in dense matter and weak magnetic field and in low
density matter and strong magnetic field. Then we demonstrate that
the relativistic quantum mechanics approach is consistent with the
previously used formalism and obtain the corrections to the
standard effective Hamiltonian. We show that the correction
obtained results in the appearance of the new resonance in
neutrino oscillations.

In Sec.~\ref{APPLICATIONS} we discuss a possible application of
our results to the conversion of right-handed electron neutrinos,
which can be created in a supernova explosion~\cite{Not88}, into a
quasi-degenerate in mass sterile neutrino when particles propagate
in an expanding envelope of a supernova and interact with its
magnetic field. The importance of other factors, which can also
influence the neutrino oscillations process, is considered in
Sec.~\ref{ANALYSIS}. Finally, in Sec.~\ref{CONCLUSION}, we
summarize our results.

\section{Relativistic quantum mechanics description\label{DENSEMATTSTRONGBQFT}}


Let us study the time evolution of the system of two mixed flavor
neutrinos $(\nu_\alpha,\nu_\beta)$ in matter and in an external
electromagnetic field $F_{\mu\nu}=(\mathbf{E},\mathbf{B})$. The
indexes $\alpha$ and $\beta$ can stand for any of the neutrino
flavors $e$, $\mu$ or $\tau$. The interaction with background
matter can be represented in terms of the external axial vector
fields $f^\mu_{\lambda\lambda'}$. The Lagrangian for our system
has the form,
\begin{align}\label{LagrmattBnu}
  \mathcal{L} = & \sum_{\lambda=\alpha,\beta}
  \bar{\nu}_\lambda
  \mathrm{i}\gamma^\mu\partial_\mu
  \nu_{\lambda} -
  \sum_{\lambda\lambda'=\alpha,\beta}
  \bar{\nu}_\lambda
  \Big(
    m_{\lambda\lambda'}+
    \gamma_\mu^\mathrm{L} f^\mu_{\lambda\lambda'}
    \notag
    \\
    & +
    \frac{1}{2}
    M_{\lambda\lambda'}
    \sigma_{\mu\nu} F^{\mu\nu}
  \Big)
  \nu_{\lambda'},
\end{align}
where $(m_{\lambda\lambda'})$ is the nondiagonal mass matrix and
$(M_{\lambda\lambda'})$ is the matrix of the neutrino magnetic
moments. Note that in general case the matrices
$(m_{\lambda\lambda'})$ and $(M_{\lambda\lambda'})$ are
independent, i.e. the diagonal form of the matrix
$(m_{\lambda\lambda'})$ in a certain basis does not imply that of
the matrix $(M_{\lambda\lambda'})$.

In the case of the standard model neutrino interactions the matrix
$(f^\mu_{\lambda\lambda'})$ is diagonal: $f^\mu_{\lambda\lambda'}
= f^\mu_\lambda \delta_{\lambda\lambda'}$, where $f^\mu_\lambda =
(f^0_\lambda, \mathbf{f}_\lambda)$. The possible nondiagonal
elements of the matrix $f^\mu_{\lambda\lambda'}$, with $\lambda
\neq \lambda'$, correspond to the non-standard neutrino
interactions with matter~\cite{Dvo08JPCS}. In the following we
will study neutrino oscillations in non-moving and unpolarized
matter with $\mathbf{f}_\lambda = 0$. The explicit form of the
zero-th component $f^0_\lambda \equiv f_\lambda$ of the four
vector $f^\mu_\lambda$ for the electoneutral medium consisting of
electrons, protons and neutrons can be found in
Ref.~\cite{DvoStu02}.
In the case of sterile neutrinos $f_{\nu_s} = 0$.

To study the time evolution of the system~\eqref{LagrmattBnu} it
is necessary to formulate the initial condition for the flavor
neutrinos $\nu_\lambda$. We suppose that only one neutrino flavor,
e.g., ``$\beta$", is present initially (see
Refs.~\cite{Dvo,Dvo06,DvoMaa07,Dvo08JPCS,Dvo08JPG}), i.e.
$\nu_\alpha(\mathbf{r},0) = 0$, and $\nu_\beta(\mathbf{r},0) =
\nu_\beta^{(0)}(\mathbf{r})$,
%
%
where $\nu_\beta^{(0)}(\mathbf{r})$ is the given function. Then we
will look for the wave function $\nu_\alpha$ at $t>0$. If we
choose $\alpha=\mu$ or $\tau$ and $\beta=e$, it will correspond to
the typical situation of neutrinos emitted in the Sun: for the
given initial flux of electron neutrinos we study the presence of
other neutrino flavors at subsequent moments of time. Note that
the analogous initial condition problem for neutrino wave packets
was studied in Ref.~\cite{BerLeo04}.

To analytically study the Dirac equation in presence of external
fields we will consider the case of the coordinate independent
functions $f_\lambda$ and $F_{\mu\nu}$. The analysis of the
validity of this approximation will be made in
Sec.~\ref{ANALYSIS}. For the coordinate independent external
fields the momentum of the particles $\mathbf{p}$ is conserved.


Moreover, the additional constraint can be imposed on the wave
function $\nu_\beta^{(0)}(\mathbf{r})$,
\begin{equation}\label{inicondhel}
  P_{\pm{}} \nu_\beta^{(0)}(\mathbf{r}) =
  \nu_\beta^{(0)}(\mathbf{r}),
  \quad
  P_{\pm{}} =
  \left(
    1\pm\frac{(\bm{\Sigma}\cdot\mathbf{p})}{|\mathbf{p}|}
  \right),
\end{equation}
where $\bm{\Sigma} = -\gamma^0\gamma^5\bm{\gamma}$ are the Dirac
matrices. Eq.~\eqref{inicondhel} implies that initially neutrinos
of the flavor ``$\beta$" have a certain helicity. If we act with
the operator $P_{\mp{}}$ on the final state
$\nu_\alpha(\mathbf{r},t)$, we can study the appearance of the
opposite helicity eigenstates among neutrinos of the flavor
``$\alpha$", i.e this situation corresponds to the neutrino spin
flavor oscillations $\nu_\beta^\mathrm{L,R} \leftrightarrow
\nu_\alpha^\mathrm{R,L}$.

Note that, unlike the helicity states defined in
Eq.~\eqref{inicondhel}, the chirality of a particle is a
eigenvalue of the matrix $\gamma^5$: $\gamma^5 \psi = \varkappa
\psi$. In case of massive Dirac particles the helicity and the
chirality do not coincide~\cite{Zub04}. The helicity is a
conserving number whereas the chirality -- not, since the operator
$\gamma^5$ does not commute with a Hamiltonian.

Let us introduce the neutrino mass eigenstates $\psi_a$, $a=1,2$,
as
\begin{equation}\label{matrtrans}
  \nu_{\lambda}=\sum_{a=1,2}U_{\lambda a}\psi_a,
  \quad
  (U_{\lambda a})
  =
  \begin{pmatrix}
    \cos \theta & -\sin \theta \\
    \sin \theta & \cos \theta \
  \end{pmatrix},
\end{equation}
to diagonalize the mass matrix $(m_{\lambda\lambda'})$. In
Eq.~\eqref{matrtrans} we take into account that for the two
neutrinos system the mixing matrix $(U_{\lambda a})$ can be
parameterized with help of one vacuum mixing angle $\theta$. We
suppose that the mass eigenstates $\psi_a$ are Dirac particles
with the masses $m_a$.

The Lagrangian~\eqref{LagrmattBnu} expressed in terms of the
fields $\psi_a$ takes the form,
\begin{align}\label{LagrmattBpsi}
  \mathcal{L} = & \sum_{a=1,2}
  \bar{\psi}_a
  (\mathrm{i}\gamma^\mu\partial_\mu - m_a)
  \psi_a
  \notag
  \\
  & -
  \sum_{ab=1,2}
  \bar{\psi}_a
  \left(
    \gamma_\mu^\mathrm{L} g^\mu_{ab}+
    \frac{1}{2}\mu_{ab}
    \sigma_{\mu\nu} F^{\mu\nu}
  \right)
  \psi_b,
\end{align}
where
\begin{align}\label{mugmeb}
  (\mu_{ab}) = & U^\dag (M_{\lambda\lambda'}) U =
  \begin{pmatrix}
    \mu_1 & \mu \\
    \mu & \mu_2 \
  \end{pmatrix},
  \notag
  \\
  (g_{ab}) = & U^\dag (f_{\lambda\lambda'}) U =
  \begin{pmatrix}
    g_1 & g \\
    g & g_2 \
  \end{pmatrix},
\end{align}
are the nondiagonal matrices of neutrino magnetic moments and
neutrino interaction with matter in the mass eigenstates basis. In
Eq.~\eqref{mugmeb} we take into account that background matter is
non-moving and unpolarized, i.e. only the zero-th component
$g_{ab} \equiv g_{ab}^0$ of the four vector $g_{ab}^\mu$ survives.

To discuss the time evolution of the system~\eqref{LagrmattBpsi}
we write down the wave equations which result from
Eq.~\eqref{LagrmattBpsi},
\begin{align}\label{DireqmattB}
  \mathrm{i}\dot{\psi}_a = & \mathcal{H}_a\psi_a+V\psi_b,
  \quad
  a=1,2,
  \quad
  a \neq b,
  \notag
  \\
  \mathcal{H}_a = & (\bm{\alpha}\mathbf{p})+\beta m_a -
  \mu_a\beta\Sigma_3 B+g_a(1-\gamma^5)/2,
  \notag
  \\
  V = & -\mu\beta\Sigma_3 B+g(1-\gamma^5)/2,
\end{align}
where $\alpha = \gamma^0\bm{\gamma}$ and $\beta = \gamma^0$ are
the Dirac matrices. Here we study the neutrino motion along the
$x$-axis: $\mathbf{p}=(p,0,0)$, in only transversal magnetic
field: $\mathbf{B}=(0,0,B)$ and $\mathbf{E}=0$.

Note that we cannot directly solve the wave
equations~\eqref{DireqmattB} because of the nondiagonal
interaction $V$ which mixes different mass eigenstates. In vacuum,
i.e in the absence of external fields, when $(g_{ab}^\mu) = 0$ and
$F_{\mu\nu} = 0$, the mass eigenstates $\psi_{1,2}$ decouple and
the system~\eqref{DireqmattB} can be easily solved. Nevertheless
we can point out an exact solution of the wave equation
$\mathrm{i}\dot{\psi}_a = \mathcal{H}_a\psi_a$, for a single mass
eigenstate $\psi_a$, that exactly accounts for the influence of
the external fields.

We look for the solution of Eq.~\eqref{DireqmattB} in the
following form~\cite{Dvo,Dvo06,DvoMaa07,Dvo08JPCS,Dvo08JPG}:
\begin{align}\label{GenSolDirpsi}
  \psi_{a}(\mathbf{r},t)= &
  e^{- \mathrm{i} g_a t/2}
  \int \frac{\mathrm{d}^3 \mathbf{p}}{(2\pi)^{3/2}}
  e^{\mathrm{i} \mathbf{p} \mathbf{r}}
  \notag
  \\
  & \times
  \sum_{\zeta=\pm 1}
  \Big[
    a_a^{(\zeta)}(t)u_a^{(\zeta)}\exp{(-\mathrm{i}E_a^{(\zeta)} t)}
    \notag
    \\
    & +
    b_a^{(\zeta)}(t)v_a^{(\zeta)}\exp{(\mathrm{i}E_a^{(\zeta)} t)}
  \Big],
\end{align}
where the energy levels, which were found in Ref.~\cite{Dvo08JPG},
have the form,
\begin{equation}\label{EnergymattB}
  E^{(\zeta)}_a = \sqrt{\mathcal{M}^2_a+m^2_a+p^2-2\zeta R^2_a},
\end{equation}
where $R^2_a=\sqrt{p^2 \mathcal{M}^2_a + (\mu_a B)^2 m^2_a}$ and
$\mathcal{M}_a=\sqrt{(\mu_a B)^2 + g_a^2/4}$. The basis spinors
can be found in the limit of a small neutrino
mass~\cite{Dvo08JPG},
\begin{widetext}
\begin{align}\label{spinorsmattB}
  u^{(\zeta)}_a = &
  \frac{1}{2\sqrt{2\mathcal{M}_a(\mathcal{M}_a-\zeta g_a/2)}}
  \begin{pmatrix}
     \mu_a B+\zeta\mathcal{M}_a-g_a/2 \\
     \mu_a B-\zeta\mathcal{M}_a+g_a/2 \\
     \mu_a B-\zeta\mathcal{M}_a+g_a/2 \\
     \mu_a B+\zeta\mathcal{M}_a-g_a/2 \
  \end{pmatrix},
  \notag
  \\
  v^{(\zeta)}_a = &
  \frac{1}{2\sqrt{2\mathcal{M}_a(\mathcal{M}_a+\zeta g_a/2)}}
  \begin{pmatrix}
     \mathcal{M}_a-\zeta[\mu_a B-g_a/2] \\
     \mathcal{M}_a+\zeta[\mu_a B+g_a/2] \\
     -\mathcal{M}_a-\zeta[\mu_a B+g_a/2] \\
     -\mathcal{M}_a+\zeta[\mu_a B-g_a/2] \
  \end{pmatrix}.
\end{align}
It should be noted that the discrete quantum number $\zeta = \pm
1$ in Eqs.~\eqref{GenSolDirpsi}-\eqref{spinorsmattB} does not
correspond to the helicity quantum states.

Now our goal is to find the time dependent coefficients
$a_a^{(\zeta)}(t)$ and $b_a^{(\zeta)}(t)$. In the case of neutrino
propagation in vacuum these coefficients do not depend on time and
their values are completely defined by the initial condition. If
we put the \textit{ansatz}~\eqref{GenSolDirpsi} in the wave
equations~\eqref{DireqmattB}, we get the following ordinary
differential equations for the function $a_a^{(\zeta)}(t)$:
\begin{equation}\label{ODEDir}
  \mathrm{i}\dot{a}_a^{(\zeta)}= 
  e^{\mathrm{i}(g_a-g_b)t/2}
  \exp\left( \mathrm{i}E_a^{(\zeta)}t \right) u^{(\zeta)\dag}_a
  V
  \sum_{\zeta'=\pm 1}
  \left[
    a_b^{(\zeta')}u^{(\zeta')}_b\exp\left( -\mathrm{i}E_b^{(\zeta')} t \right)+
    b_b^{(\zeta')}v^{(\zeta')}_b\exp\left( \mathrm{i}E_b^{(\zeta')} t \right)
  \right].
\end{equation}
To obtain Eq.~\eqref{ODEDir} we use the orthonormality of the
basis spinors~\eqref{spinorsmattB}. Note that the differential
equation for the function $b_a^{(\zeta)}$ is analogous to
Eq.~\eqref{ODEDir} and thus omitted. Moreover, taking into account
the fact that $\langle u^{(\zeta)}_a | V | v^{(\zeta')}_b \rangle
= 0$, we get that the equations for $a_a^{(\zeta)}(t)$ and
$b_a^{(\zeta)}(t)$ decouple, i.e. the interaction $V$ does not mix
positive and negative energy eigenstates.



Let us rewrite Eq.~\eqref{ODEDir} in the more conventional
effective Hamiltonian form. For this purpose we introduce the
``wave function"
$\Psi^{'\mathrm{T}}=(a_1^{-{}},a_2^{-{}},a_1^{+{}},a_2^{+{}})$.
Directly from Eq.~\eqref{ODEDir} we derive the equation for
$\Psi'$,
\begin{equation}\label{QMHprimeDir}
  \mathrm{i}\frac{\mathrm{d}\Psi'}{\mathrm{d}t} = H' \Psi',
  \quad
  H' =
  \begin{pmatrix}
    0 & h_{-{}} e^{\mathrm{i} \omega_{-{}} t} &
    0 & H_{-{}} e^{\mathrm{i} \Omega_{-{}} t} \\
    h_{-{}} e^{-\mathrm{i} \omega_{-{}} t} &
    0 & H_{+{}} e^{-\mathrm{i} \Omega_{+{}} t} & 0 \\
    0 & H_{+{}} e^{\mathrm{i} \Omega_{+{}} t} &
    0 & h_{+{}} e^{\mathrm{i} \omega_{+{}} t} \\
    H_{-{}} e^{-\mathrm{i} \Omega_{-{}} t} &
    0 & h_{+{}} e^{-\mathrm{i} \omega_{+{}} t} & 0 \
  \end{pmatrix},
\end{equation}
where $h_{\pm{}} = \langle u^{\pm{}}_a | V | u^{\pm{}}_b \rangle$,
$H_{\pm{}} = \langle u^{\pm{}}_1 | V | u^{\mp{}}_2 \rangle$,
%
%
$\omega_{\mp{}}=E_1^{\mp{}}-E_2^{\mp{}}+(g_1-g_2)/2$, and
$\Omega_{\mp{}}=E_1^{\mp{}}-E_2^{\pm{}}+(g_1-g_2)/2$. Note that we
do not give here the explicit form of the scalar products
$h_{\pm{}}$ and $H_{\pm{}}$ in order not to encumber the text.

Instead of $\Psi'$ it is more convenient to use the transformed
``wave function" $\Psi$ defined by $\Psi' = \mathcal{U}\Psi$,
where $\mathcal{U} = \mathrm{diag} \left\{
e^{\mathrm{i}(\Omega+\omega_{-{}})t/2},
e^{\mathrm{i}(\Omega-\omega_{-{}})t/2},
e^{-\mathrm{i}(\Omega-\omega_{+{}})t/2},
e^{-\mathrm{i}(\Omega+\omega_{+{}})t/2} \right\}$
%
%
and $\Omega=(\Omega_{-{}}-\Omega_{+{}})/2$. Using the property
$\omega_{+{}}+\omega_{-{}}=\Omega_{+{}}+\Omega_{-{}}$, we arrive
to the new Schr\"{o}dinger equation for the ``wave function"
$\Psi$,
\begin{equation}\label{Schr2QFT}
  \mathrm{i}\frac{\mathrm{d}\Psi}{\mathrm{d}t} = 
  H \Psi,
  \quad
  H =
  \mathcal{U}^\dag H' \mathcal{U} -
  \mathrm{i} \mathcal{U}^\dag \dot{\mathcal{U}} =
  \begin{pmatrix}
    (\Omega+\omega_{-{}})/2 & h_{-{}} & 0 & H_{-{}} \\
    h_{-{}} & (\Omega-\omega_{-{}})/2 & H_{+{}} & 0 \\
    0 & H_{+{}} & -(\Omega-\omega_{+{}})/2 & h_{+{}} \\
    H_{-{}} & 0 & h_{+{}} & -(\Omega+\omega_{+{}})/2 \
  \end{pmatrix}.
\end{equation}
Despite initially we used the analog of the perturbation theory to
analyze the influence of the potential $V$ on the dynamics of the
system~\eqref{DireqmattB}, the contribution of this potential is
exactly taken into account in Eq.~\eqref{Schr2QFT}.

As we mentioned above, the quantum number $\zeta$ does not
correspond to a definite helicity eigenstate. Thus the initial
condition, which we should add to Eq.~\eqref{Schr2QFT} depends on
the neutrino oscillations channel.
%
Suppose that one has found the solution of Eq.~\eqref{Schr2QFT} as
$\Psi^\mathrm{T}(t)=(\psi_1,\psi_2,\psi_3,\psi_4)$. Then the
transition probability for
$\nu_\beta^\mathrm{-{}}\to\nu_\alpha^\mathrm{+{}}$ oscillations
channel can be found as
\begin{align}\label{PtrsfoQFT}
  P_{\nu_\beta^\mathrm{-{}}\to\nu_\alpha^\mathrm{+{}}}(t)= &
  \frac{1}{2}
  \Bigg\{
    \frac{\mu_1 B \cos\theta}{\sqrt{\mathcal{M}_1}}
    \left[
      \frac{\psi_1(t)}{\sqrt{\mathcal{M}_1+g_1/2}}+
      \frac{\psi_3(t)}{\sqrt{\mathcal{M}_1-g_1/2}}
    \right]
    \notag
    \\
    & -
    \frac{\mu_2 B \sin\theta}{\sqrt{\mathcal{M}_2}}
    \left[
      \frac{\psi_2(t)}{\sqrt{\mathcal{M}_2+g_2/2}}
      +
      \frac{\psi_4(t)}{\sqrt{\mathcal{M}_2-g_2/2}}
    \right]
  \Bigg\}^2.
\end{align}
To obtain Eq.~\eqref{PtrsfoQFT} for simplicity we suppose that
initially we have rather broad (in space) wave packet,
corresponding to the initial condition
$\nu_\beta^{(0)}(\mathbf{r}) \sim
e^{\mathrm{i}\mathbf{p}\mathbf{r}}$.

Eqs.~\eqref{Schr2QFT} and~\eqref{PtrsfoQFT} look completely new.
However in Sec.~\ref{QM} we will show that the relativistic
quantum mechanics approach to the description of neutrino spin
flavor oscillations is consistent with the conventional quantum
mechanical approach to this problem.
\end{widetext}

\section{Quantum mechanical description\label{QM}}

In this section we analyze spin-flavor oscillations of Dirac
neutrinos in frames of the standard quantum mechanical approach.
The main concept of this approach is the construction of an
effective Hamiltonian acting in the space of quantum mechanical
neutrino ``wave functions". Thus the proper choice of the basis of
wave functions is as important as the correct form of the
effective Hamiltonian. In the majority of works devoted to
neutrino spin-flavor oscillations the basis of helicity
eigenstates was adopted (see, e.g., Ref.~\cite{LimMar88}). As we
will see below, this choice is justified only in the case of
relatively weak external magnetic field and dense matter. We also
consider the quantum mechanical derivation of the effective
Hamiltonian in the opposite situation of strong magnetic field and
low density matter. Finally we demonstrate the consistency of the
relativistic quantum mechanics approach, developed in
Sec.~\ref{DENSEMATTSTRONGBQFT}, to the standard quantum mechanical
treatment.

Historically the Schr\"{o}dinger equation which describes the
dynamics of the Dirac neutrinos system was formulated in the
flavor eigenstates basis~\cite{VolVysOku86,LimMar88}. Definitely
it is more convenient to use the flavor eigenstates basis since
one gets the transition probability directly from the solution of
the Sch\"{o}dinger equation without additional matrix
transformation~\eqref{matrtrans}. Nevertheless we will formulate
the dynamics of the neutrinos system in the mass eigenstates basis
since the energies are well defined only for the neutrino mass
eigenstates and one can distinguish the nature of neutrinos, i.e.
say whether neutrinos are Dirac or Majorana particles, only in
this basis.


Let us discuss the situation when a neutrino moves in sufficiently
dense matter and interacts with a weak magnetic field. The
opposite case of a strong magnetic field and a low density medium
will be considered later. If the matter density is great, then in
Eq.~\eqref{DireqmattB} the averaged interaction of the diagonal
magnetic moment with an external magnetic field is less than the
averaged diagonal interaction with background matter: $\langle
\mu_a\beta\Sigma_3 B \rangle \ll \langle g_a (1-\gamma^5)/2
\rangle$.
In this approximation it is convenient to rewrite the diagonal
part of the Hamiltonian in Eq.~\eqref{DireqmattB} as
\begin{align}
  \mathcal{H}_a \to & \mathcal{H}_a + \mathcal{V}_a,
  \quad
  \mathcal{V}_a = -\mu_a\beta\Sigma_3 B,
  \notag
  \\
  \mathcal{H}_a = & (\bm{\alpha}\mathbf{p})+\beta m_a +
  g_a(1-\gamma^5)/2,
\end{align}
to extract the small interaction $\mathcal{V}_a$ from the main
part of the diagonal Hamiltonian $\mathcal{H}_a$. We also rewrite
the nondiagonal interaction $V$ in Eq.~\eqref{DireqmattB} in the
following form: $V = V_B + V_m$, where $V_B = -\mu\beta\Sigma_3 B$
and $V_m = g(1-\gamma^5)/2$, to separate the nondiagonal magnetic
and matter interactions.

One can notice that now the helicity operator defined in
Eq.~\eqref{inicondhel} commutes with the modified Hamiltonian
$\mathcal{H}_a$. Therefore one can choose the helicity eigenstates
basis for the quantum mechanical ``wave functions" instead of the
more complete set of basis functions~\eqref{spinorsmattB}. If we
study neutrinos propagating along the $x$-axis, these basis
functions should satisfy the condition, $(1/2) (1 \pm \Sigma_1)
u_a^{\pm{}} = \pm u_a^{\pm{}}$. The explicit form of these spinors
can be found in Refs.~\cite{Dvo08JPCS,StuTer05}. Note that we do
not take into account the negative energy spinors $v_a^{(\zeta)}$
since in Sec.~\ref{DENSEMATTSTRONGBQFT} we demonstrated that the
evolution equations~\eqref{ODEDir} for $a_a^{(\zeta)}$ and
$b_a^{(\zeta)}$ decouple in matter and magnetic field.

Now we construct the effective $4 \times 4$ Hamiltonian
$H_{QM}^{(1)}$ which governs the dynamics of the ``wave function"
$\Psi_{QM}^{(1)\mathrm{T}} = (\psi_1^{-{}}, \psi_2^{-{}},
\psi_1^{+{}}, \psi_2^{+{}})$, where $\psi_a^{\pm{}}$ are the time
dependent $c$-number wave functions corresponding to a definite
helicity. The diagonal elements of $H_{QM}^{(1)}$ can be
calculated as the mean values of $\mathcal{H}_a$ over the states
with definite helicity $u^{(\zeta)}_a$, i.e. they are equal to the
energies of a neutrino moving only in background
matter~\cite{Dvo08JPCS,StuTer05},
\begin{equation}
  E_a^{-{}} = p + g_a + \frac{m_a^2}{2p} + \dots,
  \quad
  E_a^{+{}} = p + \frac{m_a^2}{2p} + \dots,
\end{equation}
where we use the limit of ultrarelativistic particles. The
nondiagonal elements of the effective Hamiltonian $H_{QM}^{(1)}$
are the mean values of the operators $\mathcal{V}_a$, $V_B$ and
$V_m$ over the same helicity eigenstates $u^{(\zeta)}_a$. Finally
we arrive to the effective Hamiltonian in the ultrarelativistic
limit,
\begin{equation}\label{effHammatt}
  H_{QM}^{(1)} =
  \begin{pmatrix}
    \Phi+g_1 & g & -\mu_1 B & -\mu B \\
    g & -\Phi+g_2 & -\mu B & -\mu_2 B \\
    -\mu_1 B & -\mu B & \Phi & 0 \\
    -\mu B & -\mu_2 B & 0 & -\Phi \
  \end{pmatrix},
\end{equation}
where $\Phi = \delta m^2/4k$ is the phase of vacuum oscillations
and $\delta m^2 = m_1^2 - m_2^2$. The Hamiltonian $H_{QM}^{(1)}$
determines the time evolution of the ``wave function"
$\Psi_{QM}^{(1)}$.

One can see that we have reproduced the standard effective
Hamiltonian proposed in Ref.~\cite{LimMar88} to study spin-flavor
oscillations of Dirac neutrinos. In our analysis it is important
that the helicity eigenstates $u_a^{(\zeta)}$ are used as the
basis functions. It is correct only if the diagonal magnetic
interaction $\langle \mu_a \beta \Sigma_3 B \rangle$ is small. The
numerous works devoted to neutrino spin-flavor oscillations (see,
e.g., the recent review~\cite{GiuStu09}) showed that the
Hamiltonian~\eqref{effHammatt} seems to be applicable to a more
general situation. Nevertheless here we show that the derivation
of this Hamiltonian presented, e.g., in Ref.~\cite{LimMar88} is
restricted to the case of small diagonal magnetic interaction.

Now we discuss the opposite case when a neutrino interacts with a
very strong magnetic field and moves in low density matter. Hence
in Eq.~\eqref{DireqmattB} the diagonal magnetic interaction is
much bigger than the diagonal interaction with background matter:
$\langle \mu_a\beta\Sigma_3 B \rangle \gg \langle g_a
(1-\gamma^5)/2 \rangle$. In this case it is also convenient to
redefine the diagonal Hamiltonian in Eq.~\eqref{DireqmattB} in the
following way:
\begin{align}
  \mathcal{H}_a \to & \mathcal{H}_a + \mathcal{V}_a,
  \quad
  \mathcal{V}_a = g_a(1-\gamma^5)/2,
  \notag
  \\
  \mathcal{H}_a = & (\bm{\alpha}\mathbf{p})+\beta m_a
  -\mu_a\beta\Sigma_3 B.
\end{align}
Note that now the Hamiltonian $\mathcal{H}_a$ does not conserve
the helicity of a particle. Thus to develop the standard quantum
mechanical approach we have to choose the proper basis in the
space of neutrino wave functions. For the basis wave functions it
is convenient to use the eigenvectors of the
operator~\cite{DvoMaa07}, $\Pi_a = m_a \Sigma_3 +
\mathrm{i}\gamma^0\gamma^5(\bm{\Sigma}\times\mathbf{p})_3 - \mu_a
B$, which characterizes the spin direction with respect to the
magnetic field. The explicit form of these spinors is given in
Ref.~\cite{DvoMaa07}.

The effective $4 \times 4$ Hamiltonian $H_{QM}^{(2)}$ acting in
the space of the quantum mechanical neutrino ``wave functions"
$\Psi_{QM}^{(2)\mathrm{T}} = (\psi_1^{-{}}, \psi_2^{-{}},
\psi_1^{+{}}, \psi_2^{+{}})$ can be constructed in a
straightforward way as in the case of the weak diagonal magnetic
interaction. Here $\psi_a^{\pm{}}$ are the $c$-number time
dependent functions representing neutrino states with a definite
spin projection on the magnetic field direction. However we can
notice that the Hamiltonian $H_{QM}^{(2)}$ can be obtained by the
similarity transformation, $H_{QM}^{(2)} = \mathcal{V}_{21}
H_{QM}^{(1)} \mathcal{V}_{21}^\mathrm{T}$, with the orthogonal
matrix $\mathcal{V}_{21} = (\gamma^5-\gamma^0)/\sqrt{2}$, with the
Dirac matrices $\gamma^0$ and $\gamma^5$ taken in the standard
representation~\cite{ItzZub80}.
%
%
It should be noted that the direct calculation of the Hamiltonian
$H_{QM}^{(2)}$ gives the same result.

Despite we demonstrated the consistency of the descriptions of
neutrino oscillations in different bases there is a conceptual
discrepancy between these cases. The quantum mechanical treatment
of neutrino evolution requires the correct choice of both the
effective Hamiltonian and the basis of ``wave functions" where
this Hamiltonian acts. The quantum mechanical ``wave functions"
$\Psi_{QM}^{(1,2)\mathrm{T}}$ are applicable to the description of
neutrino evolution for the cases of weak and strong diagonal
magnetic interaction respectively since they correspond to
different conserved quantum numbers in each case. Although for
ultrarelativistic neutrinos the description of neutrino
oscillations is identical in these bases, there can be a
difference if we calculate the corrections to the leading term.

\begin{widetext}
Now let us check the consistency between the relativistic quantum
mechanics method developed in Sec.~\ref{DENSEMATTSTRONGBQFT} and
the standard quantum mechanical description of neutrino
oscillations. We will study ultrarelativistic neutrinos. In this
approximation the energy levels~\eqref{EnergymattB} take the form,
\begin{equation}\label{Etotexp}
  E_a^{(\zeta)} = p + \frac{g_a}{2} - \zeta \mathcal{M}_a +
  \frac{m_a^2}{2p} + \zeta \frac{m_a^2 g_a^2}{8 p^2 \mathcal{M}_a} + \dots,
\end{equation}
where we keep the term $\sim m_a^2/k^2$ to examine the corrections
to the leading term.

Performing the similarity transformation of the effective
Hamiltonian $H$ in Eq.~\eqref{Schr2QFT} with the orthogonal matrix
$\mathcal{V}$ of the following form:
\begin{equation}\label{VQFTtoQM}
  \mathcal{V} =
  \frac{1}{\sqrt{2}}
  \begin{pmatrix}
    -\sqrt{1+g_1/2\mathcal{M}_1} & 0 &
    \mu_1 B/\sqrt{\mathcal{M}_1(\mathcal{M}_1+g_1/2)} & 0 \\
    0 &
    -\sqrt{1+g_2/2\mathcal{M}_2} &
    0 &
    \mu_2 B/\sqrt{\mathcal{M}_2(\mathcal{M}_2+g_2/2)} \\
    \sqrt{1-g_1/2\mathcal{M}_1} & 0 &
    \mu_1 B/\sqrt{\mathcal{M}_1(\mathcal{M}_1-g_1/2)} & 0 \\
    0 &
    \sqrt{1-g_2/2\mathcal{M}_2} &
    0 &
    \mu_2 B/\sqrt{\mathcal{M}_2(\mathcal{M}_2-g_2/2)} \
  \end{pmatrix},
\end{equation}
we can see that the Hamiltonian $H$ transforms to
$\mathcal{V}^\mathrm{T} H \mathcal{V} \approx H_{QM} + \delta H$,
where $H_{QM} = H_{QM}^{(1)} - \mathrm{tr}(H_{QM}^{(1)})/4 \cdot
I$, $I$ is the $4 \times 4$ unit matrix,
%
%
and
\begin{equation}\label{deltaH}
  \delta H =
  \frac{1}{16 k^2}
  \mathrm{diag}
  \left(
    -m_1^2\frac{g_1^3}{\mathcal{M}_1^2},
    -m_2^2\frac{g_2^3}{\mathcal{M}_2^2},
    m_1^2\frac{g_1^3}{\mathcal{M}_1^2},
    m_2^2\frac{g_2^3}{\mathcal{M}_2^2}
  \right),
\end{equation}
\end{widetext}
is the correction to the standard effective Hamiltonian. It should
be noted that the transformation matrix $\mathcal{V}$ in
Eq.~\eqref{VQFTtoQM} depends on the magnetic field strength and
the matter density, whereas the matrix $\mathcal{V}_{21}$ is
external field independent.

The effective Hamiltonian $H_{QM}$ is equivalent to $H_{QM}^{(1)}$
in Eq.~\eqref{effHammatt} since the unit matrix does not change
the particles dynamics. If we omit the correction~\eqref{deltaH},
that is valid for ultrarelativistic neutrinos, we can see that the
relativistic quantum mechanics approach is equivalent to the
standard quantum mechanical method. The correction $\delta H$
results from the fact that in Sec.~\ref{DENSEMATTSTRONGBQFT} we
use the correct energy levels~\eqref{EnergymattB} and the wave
functions~\eqref{spinorsmattB} for a neutrino moving in dense
matter and strong magnetic field.

Note that in Eq.~\eqref{deltaH} we keep only the diagonal
corrections to the effective Hamiltonian~\eqref{effHammatt} since
the appearance of a resonance in neutrino oscillations is
sensitive to the diagonal elements a Hamiltonian. We should remind
that the expressions for the basis spinors~\eqref{spinorsmattB}
were obtained in the approximation of small masses of neutrinos,
whereas in the expansion of the energy levels~\eqref{Etotexp} we
keep terms up to $\sim m_a^2/k^2$. However using
Eqs.~\eqref{Schr2QFT} and~\eqref{VQFTtoQM} we get that the
additional small contributions $\sim m_a^2/k^2$ from the basis
spinors are washed out in diagonal entries in Eq.~\eqref{deltaH}.

\section{Applications\label{APPLICATIONS}}

In this section we study the application of the obtained
Hamiltonians~\eqref{effHammatt} and~\eqref{deltaH} to the
description of oscillations between electron and hypothetical
sterile neutrinos in an expanding envelope of a supernova. The
existence of sterile neutrinos closely degenerate in mass with
electron, muon-{} or tau-neutrinos was recently discussed in
Ref.~\cite{Ker03} in connection to solar and supernova neutrinos.
The mass squared differences considered in these publications were
in the following range: $10^{-19}\thinspace\mathrm{eV}^2 < \delta
m^2 < 10^{-12}\thinspace\mathrm{eV}^2$.

On the contrary, the mixing angles of these additional neutrinos
cannot be well constraint. Therefore can assume that vacuum mixing
angle is small, $\theta \ll 1$. In this case mixing matrix between
mass $\psi_a$, $a=1,\dots,4$, and flavor $\nu_\lambda$,
$\lambda=e,\mu,\tau,s$, neutrino eigenstates has the form $U_{4}
\approx \mathrm{diag}(U_{3},1)$, where $U_{3}$ is the mixing
matrix of the three Dirac neutrinos system (see
Eq.~\eqref{matrtrans} and Ref.~\cite{GiuKim07}). It should be
noted that it is very difficult to detect additional neutrino
flavor if it is weakly mixed with active neutrinos and $\delta
m^2$ is small. Such a neutrino can be revealed through spin-flavor
oscillations only if it has a transition magnetic moment.


Besides the huge amount of left-handed neutrinos from a supernova,
a smaller flux of right-handed particles is
predicted~\cite{Not88}. These right-handed neutrinos can be
created in the following reaction: $\nu_\mathrm{-{}}+(e^{-{}}, p,
N) \to \nu_\mathrm{+{}}+(e^{-{}}, p, N)$, with electrons
$e^{-{}}$, protons $p$, and nuclei $N$ in the dense matter of a
protoneutron star. For the neutrino spin-flip in matter to happen
within one generation $\nu_\alpha^\mathrm{-{}} \to
\nu_\alpha^\mathrm{+{}}$, a neutrino should be the Dirac particle
with a nonzero diagonal magnetic moment. Left-polarized supernova
neutrinos are strongly degenerate and can occupy energy levels
above the Fermi surface. Right-polarized neutrinos created in the
spin-flip reactions have the energy of the order of the
left-polarized neutrinos. The detailed analysis of
Ref.~\cite{Not88} shows that the energy of these particles is in
the $E_\nu = (100-200)\thinspace\mathrm{MeV}$ range.

The generation of electron neutrinos with right-handed
polarization can provide additional supernova cooling since they
do not interact with background matter and thus freely carry away
the supernova energy. Moreover these particles can be potentially
detected in a terrestrial detector due to their spin precession,
back to left-handed states, in the galactic magnetic field.
However, if we point out an additional neutrino oscillations
channel, which contributes to the right-handed neutrinos dynamics,
these particles are unlikely to be observed.

Let us study the appearance of a resonance in $\nu_e^\mathrm{+{}}
\leftrightarrow \nu_s^\mathrm{-{}}$ oscillations channel. It is
known that a resonance in a certain channel of neutrino
oscillations can appear if the difference between two diagonal
elements in the effective Hamiltonian is small~\cite{AkhLanSci97}.
Using Eqs.~\eqref{effHammatt} and~\eqref{deltaH} in the
approximation of $\theta \ll 1$, as well as the results of
Ref.~\cite{DvoStu02} we obtain the resonance condition for these
oscillations as,
\begin{align}\label{rescond2}
  \delta m^2 \approx & 5.0 \times 10^{-17}\thinspace\mathrm{eV}^2
  \times
  (3Y_e-1)
  \notag
  \\
  & \times
  \left(
    \frac{\rho}{10^6\thinspace\mathrm{g/cm}^3}
  \right)
  \left(
    \frac{E_\nu}{100\thinspace\mathrm{MeV}}
  \right)^{-1}
  \left(
    \frac{m_{\nu_e}}{1\thinspace\mathrm{eV}}
  \right)^2,
\end{align}
where $Y_e=n_e/(n_e+n_n)$ is the electrons fraction, $\rho$ is the
mass density of background matter, and $m_{\nu_e}$ is the absolute
mass of an electron neutrino. Since $\delta m^2$ is supposed to be
small, we can set $m_\nu$ to be equal to either $m_1$ or the mass
the additional neutrino mass eigenstate appearing due to the
presence of a sterile neutrino. In Eq.~\eqref{rescond2} we suppose
that matter is electroneutral and the diagonal magnetic moment of
an electron neutrino is small: $\mu_{\nu_e} B \ll f_e$.

Suppose that the flux of right-handed electron neutrinos is
crossing an expanding envelope of a supernova. A shock wave is
usually formed in the envelope~\cite{Tom05}. Approximately
$1\thinspace\mathrm{s}$ after the core collapse, the matter
density in the shock wave region $L \sim
10^8\thinspace\mathrm{cm}$ can be up to
$10^6\thinspace\mathrm{g/cm}^3$. We can also suppose that the
matter density is approximately constant inside the shock wave.
From Eq.~\eqref{rescond2} we can see that a resonance in neutrino
oscillations happens if $\rho \approx
10^6\thinspace\mathrm{g/cm}^3$, $Y_e > 1/3$, $E_\nu \sim
100\thinspace\mathrm{MeV}$, $m_{\nu_e} \approx
0.2\thinspace\mathrm{eV}$, and $\delta m^2 \approx 2\times
10^{-18}\thinspace\mathrm{eV}^2$, that is consistent with the
modern cosmological limits on the absolute neutrino
mass~\cite{Kra05}, the mentioned above estimates of the mass
squared differences with a sterile neutrino and the energy of
right-handed neutrinos, as well as to the parameters of a shock
wave.

Besides the fulfillment of the resonance
condition~\eqref{rescond2}, to have the significant
$\nu_e^\mathrm{+{}} \leftrightarrow \nu_s^\mathrm{-{}}$
transitions rate the oscillations length should be comparable to
the the shock wave size $L$. We can express this condition in the
following form:
\begin{equation}\label{rescond3}
  B \approx 5.3 \times 10^7\thinspace\mathrm{G}
  \times
  \left(
    \frac{\mu}{10^{-12}\thinspace\mu_\mathrm{B}}
  \right)^{-1}
  \left(
    \frac{L}{10^{3}\thinspace\mathrm{km}}
  \right)^{-1}.
\end{equation}
%
For the transition magnetic moment $\mu = 3 \times
10^{-12}\thinspace\mu_\mathrm{B}$~\cite{Raf90} and $L \sim
10^{8}\thinspace\mathrm{cm}$ (see above), we get that $B \sim
10^7\thinspace\mathrm{G}$. Supposing that the magnetic field of a
protoneutron star depends on the distance as $B_\mathrm{dip}(r) =
B_0 (R/r)^3$, where $R = 10\thinspace\mathrm{km}$ is the typical
protoneutron star radius and $B_0 = 10^{13}\thinspace\mathrm{G}$
is the magnetic field on the surface of a protoneutron star, we
get that at $r = 10^8\thinspace\mathrm{cm}$ the magnetic field
reaches $10^7\thinspace\mathrm{G}$, which is consistent with the
estimates of Eq.~\eqref{rescond3}. Note that magnetic fields in a
supernova explosion can be even higher than
$10^{13}\thinspace\mathrm{G}$ and reaches the values of $\sim
10^{16}\thinspace\mathrm{G}$~\cite{Aki03}.

\begin{figure*}
  \centering
  \includegraphics[scale=1]{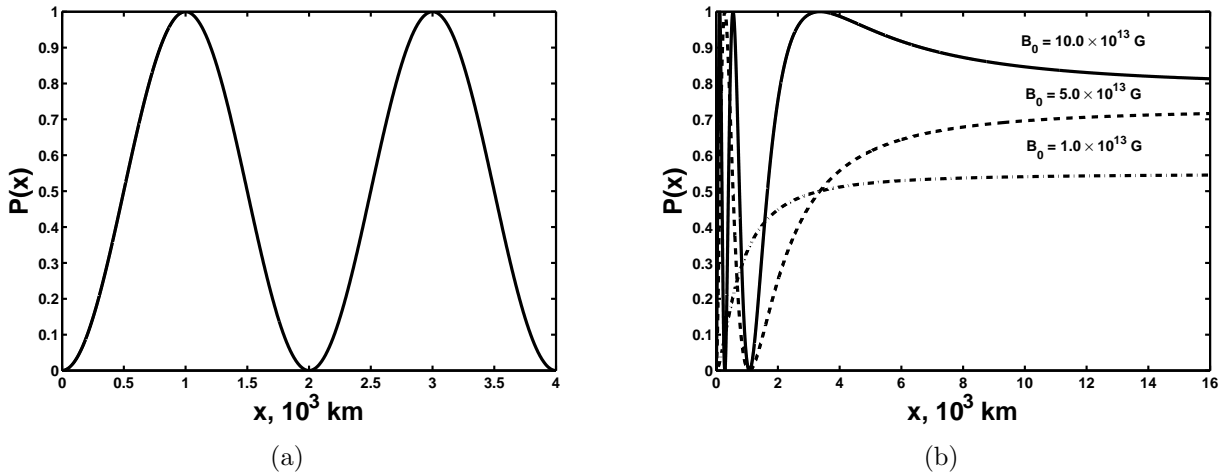}
  \caption{\label{Ptrnum}
  (a) The numerical transition probability versus the distance traveled by
  a neutrino beam when both resonance conditions, Eqs.~\eqref{rescond2}
  and~\eqref{rescond3}, are satisfied. The parameters of the neutrino system are the
  following: $\mu = 3 \times 10^{-12}\thinspace\mu_\mathrm{B}$,
  $\mu_{\nu_e} = 10^{-13}\thinspace\mu_\mathrm{B}$, $\mu_{\nu_s} = 0$,
  $E = 100\thinspace\mathrm{MeV}$,
  $m_{\nu_e} = 1\thinspace\mathrm{eV}$ and
  $\delta m^2 = 5 \times 10^{-17}\thinspace\mathrm{eV}^2$. Neutrinos move in
  background matter with $\rho = 10^6\thinspace\mathrm{g/cm}^3$ and
  interact with the magnetic field $B = 10^7\thinspace\mathrm{G}$.
  (b) The numerical transition probability as a function of the distance in the case of
  the coordinate dependent magnetic field $B_\mathrm{dip}(r)$ for various values of
  $B_0$. The distance $x$ is measured from the inner radius of an envelope
  $R_\mathrm{env} = 10^8\thinspace\text{cm}$. The parameters of the neutrino system are
  the same as in the panel (a).}
\end{figure*}

In Fig.~\ref{Ptrnum}(a)
we present the numerical solution of the Schr\"{o}dinger equation
with the Hamiltonians~\eqref{effHammatt} and~\eqref{deltaH} taking
into account the diagonal magnetic moment of an electron neutrino
which was neglected while obtaining the resonance conditions since
we assumed that $\mu_a \ll \mu$. It is necessary to remind that in
Ref.~\cite{Not88} it was found that to get a significant
$\nu_e^\mathrm{+{}}$ luminosity $\sim
10^{50}\thinspace\mathrm{erg/s}$ the diagonal magnetic moment
should be $\mu_{\nu_e} = 10^{-13}\thinspace\mu_\mathrm{B}$. We
suppose that both Eqs.~\eqref{rescond2} and~\eqref{rescond3} are
satisfied.
One can see in Fig.~\ref{Ptrnum}(a) that diagonal magnetic moments
do not significantly influence the dynamics of spin-flavor
oscillations since the numerical transition probability
practically coincide with the approximate analytical expression
$P(x) = \sin^2 (\mu B x)$ obtained from Eqs.~\eqref{effHammatt},
\eqref{deltaH}, \eqref{rescond2}, and~\eqref{rescond3} in the
limit of small $\mu_{\nu_e}$.

We plot Fig.~\ref{Ptrnum}(a) in the approximation of constant
external fields. Despite the matter density decreases as $1/r^3$
inside the envelope~\cite{Kac02}, we can suppose that it is
approximately constant in the shock wave region. However the
approximation of the constant magnetic field is quite rough. In
Fig.~\ref{Ptrnum}(b) we present the solution of the exact
Schr\"{o}dinger equation for the coordinate dependent magnetic
field $B_\mathrm{dip}(r)$, for various values of the magnetic
fields $B_0$ at the neutrinosphere. We again suppose that the
resonance condition~\eqref{rescond2} is fulfilled.

Note that in Sec.~\ref{DENSEMATTSTRONGBQFT} we assumed that the
magnetic field is coordinate independent. However, if we suggest
that the length of the spatial variation of a magnetic field is
much bigger than the typical size of a neutrino wave packet (see
also Sec.~\ref{ANALYSIS}), we can neglect it in the analysis of
the neutrino wave equation~\eqref{DireqmattB}. Thus, if this
condition is satisfied, the magnetic field is supposed to be
spatially constant in deriving main Eqs.~\eqref{Schr2QFT},
\eqref{effHammatt}, and~\eqref{deltaH} in
Secs.~\ref{DENSEMATTSTRONGBQFT} and~\ref{QM}. Nevertheless we take
into account the magnetic field variation while solving the
effective Schr\"{o}dinger equation with the
Hamiltonians~\eqref{effHammatt} and~\eqref{deltaH}. This
requirement for spatial variations of magnetic fields can be
characterized as the microscopic adiabaticity condition in
contrast to the macroscopic adiabaticity typically used in the
analysis of neutrino oscillations (see, e.g.,
Ref.~\cite{MohPal04}).

One can see in Fig.~\ref{Ptrnum}(b) that at relatively weak
magnetic fields $B_0 \sim 10^{13}\thinspace\text{G}$ the
transition probability is a strictly increasing function reaching
the asymptotic value $\sim 0.5$ and never is equal to one. This
magnetic field corresponds to $B(r = R_\textrm{env}) =
10^7\thinspace\text{G}$, where $R_\textrm{env} =
10^8\thinspace\text{cm}$ is the internal radius of an expanding
envelope. We should remind that oscillations in a constant
magnetic field of such a strength are at resonance, cf.
Eq.~\eqref{rescond3}, and hence the transition probability can
reach a unit value.

We also notice that at the big distances traveled by neutrinos the
transition probability becomes constant. Indeed, using
Eqs.~\eqref{effHammatt} and~\eqref{deltaH} we get that in the
limit of the small diagonal magnetic moment of an electron
neutrino and if the resonance condition~\eqref{rescond2} is
fulfilled, at big distances the transition probability has the
form, $P(x \gg R_\textrm{env}) = \sin^2(\mu B_0 R^3/2
R_\textrm{env}^2)$. The values of the asymptotic transition
probability shown in Fig.~\ref{Ptrnum} are in agreement with this
expression. At strong magnetic fields ($B_0 =
10^{14}\thinspace\text{G}$ or $B(r = R_\textrm{env}) =
10^8\thinspace\text{G}$) the transition probability can reach big
values at the outer edge of a broad envelope, $r \sim
10^9\thinspace\text{cm}$. Note that analogous behaviour of
transition probability was found in Ref.~\cite{TotSat96} while
studying spin-flavor oscillations of Majorana neutrinos in a
supernova.

\section{Analysis of approximations\label{ANALYSIS}}


First we should remind that we use the relativistic quantum
mechanics approach, with the external fields being independent of
spatial coordinates. If external fields depend on the spatial
coordinates, Dirac wave packets theory reveals various additional
phenomena such as particles creation by the external field
inhomogeneity~\cite{ItzZub80p60}. For the approximation of the
spatially constant external fields to be valid, the typical length
scale of the external field variation $L_\mathrm{ext}$ should be
much greater than the Compton length of a
neutrino~\cite{ItzZub80p60}: $L_\mathrm{ext} \gg \lambdabar_C =
\hbar/m_\nu c$~\cite{DvoStu02}. For a neutrino with $m_\nu \sim
1\thinspace\mathrm{eV}$ this condition reads $L_\mathrm{ext} \gg
10^{-5}\thinspace\mathrm{cm}$, that is fulfilled for almost all
realistic external fields.

We should also make a remark on the accounting for the mixing
potential $V$ in Eq.~\eqref{ODEDir}. Eq.~\eqref{DireqmattB}
contains the nondiagonal term $\sim V$. The analysis of analogous
equations is typically made in frames of the perturbation
theory~\cite{LanLif02} using the expansion on powers of the mixing
potential, i.e. on the coupling constants which are proportional
to $g$ and $\mu B$. As it was mentioned in
Sec.~\ref{DENSEMATTSTRONGBQFT}, the wave equations for massive
neutrinos in vacuum decouple and the evolution of these states
depends on the initial condition only. While solving
Eq.~\eqref{ODEDir} we could have taken into account the terms
linear in $g$ and $\mu B$ as it was made in
Refs.~\cite{Dvo06,DvoMaa07}. However in subsequent calculations,
which lead to Eqs.~\eqref{QMHprimeDir} and~\eqref{Schr2QFT}, these
coefficients were accounted for exactly.

Now let us discuss other factors which can also give the
contributions, comparable with Eq.~\eqref{deltaH}, to the
effective Hamiltonian. While deriving the effective
Hamiltonian~\eqref{Schr2QFT} in Sec.~\ref{DENSEMATTSTRONGBQFT} we
supposed that the magnetic field is transverse with respect to the
neutrino motion. The effect of the longitudinal magnetic field
$B_{\parallel{}}$ on neutrino oscillations was studied in
Ref.~\cite{AkhKhl88}.
In order to neglect the longitudinal magnetic field contribution
in comparison with our correction~\eqref{deltaH}, its strength
should satisfy the condition,
\begin{equation}\label{Blong}
  \frac{B_{\parallel{}}}{B_{\perp{}}} \ll
  \frac{1}{16  k B_{\perp{}}|\mu_a m_a - \mu_b m_b|}
  \left|
    \frac{m_a^2 g_a^3}{\mathcal{M}_a^2}-
    \frac{m_b^2 g_b^3}{\mathcal{M}_b^2}
  \right|,
\end{equation}
where $B_{\perp{}}$ is the transverse component of the magnetic
field.
%
%
The condition~\eqref{Blong} is satisfied for neutrinos emitted
inside the solid angle near the equatorial plane with the spread
$2\vartheta \sim 2 B_{\parallel{}}/B_{\perp{}}$. Assuming the
radially symmetric neutrino emission we find that about 2\% of the
total neutrino flux is affected by the new
resonance~\eqref{rescond2}, i.e. the influence of the longitudinal
magnetic field is negligible for oscillations of such particles.

The next important approximation made in the deviation of
Eq.~\eqref{Schr2QFT} was the assumption of negligible polarization
of matter which can be not true if we study rather strong magnetic
fields. The effect of matter polarization on neutrino oscillations
was previously discussed in Refs.~\cite{DvoStu02,Nun97,LobStu01}.
It was found that in the leading order in $m_a/k$ matter
polarization produces the following contribution to the diagonal
entry of the effective Hamiltonian, corresponding to right-handed
neutrinos: $g_a (\bm{\lambda}_f \bm{\beta}_\nu) (m_a/k)$, where
$\bm{\beta}_\nu$ is the neutrino velocity and $\bm{\lambda}_f$ is
the mean polarization vector of background fermions. Note that in
polarized matter the effective energy of left-handed neutrinos
also changes. However this process does not contribute to the
considered oscillations channel $\nu_e^\mathrm{+{}}
\leftrightarrow \nu_s^\mathrm{-{}}$.

It is clear that one should take into account only the
polarization of electrons since nucleons are much heavier. Using
the results of Refs.~\cite{Nun97,Sum05} we obtain that in the
shock wave region electrons are relativistic and weakly
degenerate. Finally we get that the new correction to the
effective Hamiltonian~\eqref{deltaH} becomes bigger than the
contribution of matter polarization to the effective potential of
right-handed neutrinos if the electron temperature exceeds
$10\thinspace\mathrm{MeV}$ which is comparable with the mean
temperature in an expanding envelope~\cite{Sum05}.

The presence of relatively big Dirac neutrino magnetic moments
implies the existence of new interactions, beyond the standard
model, which electromagnetically couple left-{} and right-handed
neutrinos. It is possible that these new interactions also
contribute to the effective potential of the right-handed neutrino
interaction with background matter. Despite this additional
effective potential is likely to be small, one should evaluate it
and compare with the correction~\eqref{deltaH}.

The most generic $\mathrm{SU}(2)_L \times \mathrm{U}(1)_Y$ gauge
invariant and remormalizable interaction which produces Dirac
neutrino magnetic moment was discussed in Ref.~\cite{Bel05}. The
effective Lagrangian of this interaction involves the dimension
$n=6$ operators $\mathcal{O}_j$, $\mathcal{L}_\mathrm{eff} =
(1/\Lambda^2) \sum_j C_j \mathcal{O}_j + \mathrm{h.c.}$, where
$\Lambda \sim 1\thinspace\mathrm{TeV}$ is scale of the new
physics, $C_j$ are the effective operator coupling constants and
sum spans all the operators of the given dimension.

One of the operators $\mathcal{O}_j$ also contributes to the
effective potential of a right-handed neutrino in matter,
$\mathcal{O} = \kappa \bar{L} \tau_a \tilde{\phi} \sigma^{\mu\nu}
\nu_\mathrm{R} W_{\mu\nu}^a$, where $\kappa$ is the coupling
constant, $\tau_a$ are Pauli matrices,
$L^\mathrm{T} =
(\nu_\mathrm{L},e_\mathrm{L})$ is the $\mathrm{SU}(2)_L$
isodoublet, $\tilde{\phi} = \mathrm{i} \tau_2 \phi^{*{}}$, with
$\phi$ being a Higgs field, and $W_{\mu\nu}^a =
\partial_\mu W_\nu^a - \partial_\nu W_\mu^a - \kappa \epsilon_{abc} W_\mu^b
W_\nu^c$ is the $\mathrm{SU}(2)_L$ field strength tensor. Assuming
the spontaneous symmetry breaking at the electroweak scale,
$\phi^\mathrm{T} \to (0, v/\sqrt{2})$, we can rewrite the
contribution of the operator $\mathcal{O}$ to the effective
Lagrangian in the form,
\begin{equation}
  \mathcal{L}_\mathrm{eff} = \frac{C \kappa v}{\sqrt{2}}
  \bar{e}_{L} \sigma^{\mu\nu} \nu_{R} (W_{\mu\nu}^1 - {i} W_{\mu\nu}^2) +
  \mathrm{h.c.},
\end{equation}
which implies that a process $e^{-{}}+\nu_\mathrm{+{}} \to
e^{-{}}+\nu_\mathrm{-{}}$ should happen in background matter.

Using the results of Ref.~\cite{Bel05} we can evaluate the
contribution of the new interactions to the effective
Hamiltonian~\eqref{effHammatt} as
\begin{equation}
  \delta V_{R} \sim V_\mathrm{sm}
  \frac{|\kappa|^2}{G_\mathrm{F} M_W^2}
  \left(
    \frac{\mu_\nu}{\mu_\mathrm{B}}
  \right)^2
  \left(
    \frac{E_\nu}{m_e}
  \right)^2,
\end{equation}
where $V_{sm} \sim G_\mathrm{F} n_e$ is the standard model
effective potential, $G_\mathrm{F}$ is the Fermi constant, and
$M_W$ is the $W$ boson mass. Taking $\mu_\nu \sim 10^{-12}
\mu_\mathrm{B}$, $E_\nu \sim 100\thinspace\mathrm{MeV}$ and $m_\nu
\sim 0.2\thinspace\mathrm{eV}$ (see Sec.~\ref{APPLICATIONS}) we
can get that the ratio of the correction to the effective
potential $\delta V_R$ and new correction~\eqref{deltaH} is $\sim
10^{-2}$. It means that the influence of new interactions, which
generate neutrino magnetic moments, are not important for neutrino
spin-flavor oscillations.

The constraint on the Dirac neutrino magnetic moment obtained in
Ref.~\cite{Bel05} is $10^{-14} \mu_\mathrm{B}$. Nevertheless in
Sec.~\ref{APPLICATIONS} we used the magnetic moments in the
$(10^{-13} - 10^{-12})\mu_\mathrm{B}$ range since analogous
constrains on the Dirac neutrino magnetic moments were obtained in
Refs.~\cite{Not88,Raf90} on the basis of the analysis of
astrophysical data.

\section{Conclusion\label{CONCLUSION}}

In this paper we have described neutrino spin-flavor oscillations
in dense matter and strong magnetic field in the frame of
relativistic quantum mechanics. The advantage of this formalism,
compared to the commonly used quantum mechanical approach, is that
one can exactly take into account the neutrino properties like
initial momentum, $\delta m^2$, mixing angles and magnetic
moments, as well as the matter density and the strength of the
magnetic field since we used the exact solutions of the Dirac
equation for massive neutrinos in presence of external fields.

In Sec.~\ref{DENSEMATTSTRONGBQFT} it was demonstrated that the
initial condition problem for the system of two mixed flavor
neutrinos, each of them represented as four-component Dirac
spinors, can be reduced to a Schr\"{o}dinger like equation [see
Eq.~\eqref{Schr2QFT}]. In Sec.~\ref{QM} it was shown that the
Hamiltonian of this evolution equation formally coincides with the
previously proposed~\cite{LimMar88} effective
Hamiltonian~\eqref{effHammatt}.

It should be, however, noted that the dynamics of neutrino
spin-flavor oscillation in matter and magnetic field in frames of
the quantum mechanical description is defined by both the
effective Hamiltonian and the correct basis of the neutrinos wave
functions. In Sec.~\ref{QM} we showed that the choice of the
helicity eigenstates basis is justified only in the case of the
weak magnetic field limit. This kind of basis was used in the
previous treatment of neutrino spin-flavor
oscillations~\cite{LimMar88}. The opposite situation of the strong
magnetic field and low density matter was also discussed in
Sec.~\ref{QM}. In this limit the correct basis consists of the
eigenfunctions of the operator $\Pi_a$.

Besides the demonstration of the consistency of our results with
the standard approach for the description of neutrino spin-flavor
oscillations, we found the correction~\eqref{deltaH} to the
commonly used effective Hamiltonian~\eqref{effHammatt} which is
usually omitted~\cite{LimMar88}. It was possible to obtain this
correction in the explicit form since we used the energy
levels~\eqref{EnergymattB} and basis spinors~\eqref{spinorsmattB}
which exactly account for external fields.

In Sec.~\ref{APPLICATIONS} we discussed the realistic situation
when the correction~\eqref{deltaH} is important. We considered
$\nu_e^\mathrm{+{}} \leftrightarrow \nu_s^\mathrm{-{}}$
oscillations channel, where $\nu_s$ is the additional sterile
neutrino almost degenerate in mass with other neutrino states. The
right-handed electron neutrinos were supposed to be produced
during the supernova explosion due to the scattering of
left-handed neutrinos with the non-zero diagonal magnetic moment
on background fermions~\cite{Not88}. The flux of these
right-handed neutrinos was taken to propagate through the
expanding envelope and interact with an external magnetic field
due to the presence of the non-zero transition magnetic moment.

We found that the new resonance in neutrino spin-flavor
oscillations can appear if the strength of the magnetic field and
the matter density are close to the values recently discussed in
Refs.~\cite{Aki03,Tom05}.
Note that new resonance condition~\eqref{rescond2} depends on the
absolute value of the neutrino mass. Thus the observation of this
new resonance would provide the information about the absolute
scale of the neutrino masses. Although $\nu_e^\mathrm{+{}}
\leftrightarrow \nu_s^\mathrm{-{}}$ oscillations do not change the
dynamics of a supernova explosion the appearance of this
additional resonance channel of neutrino spin-flavor oscillations
makes impossible a terrestrial observation of the right-handed
supernova neutrinos proposed in Ref.~\cite{Not88}.

To analyze the dynamics of the Schr\"{o}dinger equation with exact
Hamiltonians~\eqref{effHammatt} and~\eqref{deltaH} in
Sec.~\ref{APPLICATIONS} we presented the numerical transition
probability which accounts for all magnetic moments and is built
for various magnetic fields configurations. In particular we
analyze the constant magnetic field and more realistic coordinate
dependent magnetic field of a magnetic dipole $B_\mathrm{dip}(r)$.
It was revealed that diagonal magnetic moment of an electron
neutrino does not significantly influence the transition
probability. Then we found that at rather strong magnetic field
strength the transition probability can reach big values when a
neutrino leaves an expanding envelope.

In Sec.~\ref{ANALYSIS} we considered other factors which can be
comparable with the correction obtained~\eqref{deltaH}. We
analyzed the contributions of the longitudinal magnetic field and
matter polarization which were omitted in the derivation of the
effective Hamiltonian~\eqref{Schr2QFT}.
In particular we have found that a longitudinal magnetic field is
not important for neutrinos emitted near the equator of a star.
The matter
polarization does not influence the dynamics of neutrino
oscillations if the temperature of background electrons is higher
than a few MeV. We also evaluated the possible contributions of
the new interactions, which generate neutrino magnetic
moments~\cite{Bel05}, to the effective potentials of right-handed
neutrinos. It was found that these contributions are negligible
compared to the correction~\eqref{deltaH}.

\begin{acknowledgments}
  This work has been supported by CONICYT (Chile) through Programa
  Bicentenario PSD-91-2006, by Deutscher Akademischer Austausch Dienst,
  and by FAPESP (Brazil).
  The author is thankful to
  G.~G.~Raffelt and V.~B.~Semikoz for helpful discussions.
\end{acknowledgments}

\end{document}